\documentclass{PoS}
\usepackage{epsfig,float}

\newcommand{\be}{\begin{equation}}
\newcommand{\ee}{\end{equation}}
\newcommand{\ba}{\begin{eqnarray}}
\newcommand{\ea}{\end{eqnarray}}

\newcommand{\De}{\Delta}

\title{Dynamically generated resonances from the interaction of vector mesons
with baryons}

\ShortTitle{Vector-baryon resonances}

\author{\speaker{E. Oset}\\
        Departamento de F\'{\i}sica Te\'orica and IFIC,
Centro Mixto Universidad de Valencia-CSIC,
Institutos de Investigaci\'on de Paterna, Aptdo. 22085, 46071 Valencia, Spain\\
        E-mail: \email{oset@ific.uv.es}}
	
\author{P. Gonzalez, M. J. Vicente Vacas \\
        Departamento de F\'{\i}sica Te\'orica and IFIC,
Centro Mixto Universidad de Valencia-CSIC,
Institutos de Investigaci\'on de Paterna, Aptdo. 22085, 46071 Valencia, Spain\\
       }	

\author{A. Ramos\\
       Departament d'Estructura i Constituents de la Materia, Universitat de
Barcelona\\
       }

\author{J. Vijande\\
       Departamento de Fisica Atomica Molecular y Nuclear, and IFIC, Universidad de
Valencia\\
      }

\author{S. Sarkar\\
       Variable Energy Cyclotron Centre, 1/AF, Bidhannagar, Kolkata 700064, India \\
       }

\author{Bao Xi Sun\\
    Institute of Theoretical Physics, College of Applied Sciences,
Beijing University of Technology, Beijing 100124, China \\
       }

\abstract{ We present the results of the first calculations involving the
interaction of vector mesons with baryons, by means of which one generates a
large amount of dynamically generated resonances, many of which can be
associated to known resonances, while others represent predictions for new
states.}

\FullConference{International Workshop on Effective Field Theories: from the pion to the upsilon \\
		February 2-6 2009\\
		Valencia, Spain}

\begin{document}

\section{Introduction}

The combination of the information contained in the chiral Lagrangians 
 and unitary techniques in coupled channels of mesons and baryons 
is an efficient tool to face many problems in Hadron Physics. In this way the 
interaction of the octet of
pseudoscalar mesons with the octet of stable baryons has been studied and 
leads to $J^P=1/2^-$
resonances which fit quite well the spectrum of the known low lying resonances
with these quantum numbers 
\cite{Kaiser:1995cy,angels,ollerulf,carmenjuan,hyodo}. 
Similarly the interaction
of the octet of pseudoscalar mesons with the decuplet of baryons also leads to
many resonances that can be identified with existing ones of $J^P=3/2^-$
\cite{kolodecu,sarkar}. Sometimes a new
resonance is predicted, as in the case of the $\Lambda(1405)$, where all the
chiral approaches find two close poles rather than one, for which 
experimental support is presented in \cite{magas}.  Another step forward in this
 direction has been the interpretation
of low lying $J^P=1/2^+$ states as molecular systems of two pseudoscalar mesons and one baryon
\cite{alberto,alberto2,kanchan,Jido:2008zz,KanadaEn'yo:2008wm}. 

Much work has been done using pseudoscalar
mesons as building blocks, but the consideration of vectors instead of
pseudoscalars is only beginning to be exploited. In the baryon sector the
interaction of the $\rho$ $\Delta$ has been recently addressed in
\cite{vijande}, where three degenerate $N^*$ states and three degenerate
$\Delta$ states around $1900~MeV$, with $J^P=1/2^-, 3/2^-, 5/2^-$, are found. The
underlying theory for this study is the hidden gauge formalism
\cite{hidden1,hidden2}, which deals with the interaction of vector mesons and
pseudoscalars  respecting chiral dynamics, providing the interaction of
pseudoscalars among themselves, with vector mesons, and vector mesons among
themselves. It also offers a perspective on the chiral Lagrangians as limiting
cases at low energies of vector exchange diagrams occurring in the theory. 
The extrapolation to SU(3) with the interaction of the vectors of the nonet with
the baryons of the decuplet has been done in \cite{sourav}.

 In the meson sector, the interaction of $\rho \rho$ within this formalism has
been addressed in \cite{raquel}, where it has been shown to lead  to the
dynamical generation of the $f_2(1270)$ and $f_0(1370)$  meson resonances, with
a branching ratio for the sensitive $\gamma \gamma$ decay channel in good
agreement with experiment \cite{junko}. The extrapolation to SU(3) of the work 
of \cite{raquel} has been done in \cite{gengvec}, where many resonances are
obtained, some of which can be associated to known meson states, while there are
predictions for new ones.

  In this talk we present the results of the interaction of the nonet
   of vector mesons with the decuplet of baryons \cite{sourav} and with the
   octet of baryons \cite{angelsvec}, which have been done 
 using the unitary approach in coupled
channels. The scattering amplitudes lead to poles in the
complex plane which can be associated to some well known resonances. Under the
approximation of neglecting the three momentum of the particles versus their
mass, implicit in the chiral Lagrangians, we obtain degenerate states of $J^P=1/2^-,3/2^-$ for the case of the
interaction with the octet of baryons and $J^P=1/2^-,3/2^-,5/2^- $  for the
case of the interaction with the baryons of the decuplet. This degeneracy 
seems to be followed qualitatively by the experimental spectrum, although in
some cases the spin partners have not been identified.

\section{Formalism for $VV$ interaction}

We follow the formalism of the hidden gauge interaction for vector mesons of 
\cite{hidden1,hidden2} (see also \cite{hidekoroca} for a practical set of Feynman rules). 
The Lagrangian involving the interaction of 
vector mesons amongst themselves is given by
\begin{equation}
{\cal L}_{III}=-\frac{1}{4}\langle V_{\mu \nu}V^{\mu\nu}\rangle \ ,
\label{lVV}
\end{equation}
where the symbol $\langle \rangle$ stands for the trace in the $SU(3)$ space 
and $V_{\mu\nu}$ is given by 
\begin{equation}
V_{\mu\nu}=\partial_{\mu} V_\nu -\partial_\nu V_\mu -ig[V_\mu,V_\nu]\ ,
\label{Vmunu}
\end{equation}
with  $g$ given by $g=\frac{M_V}{2f}$
with $f=93\,MeV$ the pion decay constant. The magnitude $V_\mu$ is the $SU(3)$ 
matrix of the vectors of the nonet of the $\rho$
\begin{equation}
V_\mu=\left(
\begin{array}{ccc}
\frac{\rho^0}{\sqrt{2}}+\frac{\omega}{\sqrt{2}}&\rho^+& K^{*+}\\
\rho^-& -\frac{\rho^0}{\sqrt{2}}+\frac{\omega}{\sqrt{2}}&K^{*0}\\
K^{*-}& \bar{K}^{*0}&\phi\\
\end{array}
\right)_\mu \ .
\label{Vmu}
\end{equation}

The interaction of ${\cal L}_{III}$ gives rise to a contact term coming from 
$[V_\mu,V_\nu][V_\mu,V_\nu]$
\begin{equation}
{\cal L}^{(c)}_{III}=\frac{g^2}{2}\langle V_\mu V_\nu V^\mu V^\nu-V_\nu V_\mu
V^\mu V^\nu\rangle\ ,
\label{lcont}
\end{equation}
 and on the other hand it gives rise to a three 
vector vertex from 
\begin{equation}
{\cal L}^{(3V)}_{III}=ig\langle (\partial_\mu V_\nu -\partial_\nu V_\mu) V^\mu V^\nu\rangle
\label{l3V}=ig\langle (V^\mu\partial_\nu V_\mu -\partial_\nu V_\mu
V^\mu) V^\nu\rangle
\label{l3Vsimp}\ ,
\end{equation}

In this latter case one finds an analogy with the coupling of vectors to
 pseudoscalars given in the same theory by 
 
\be
{\cal L}_{VPP}= -ig ~tr\left([
P,\partial_{\mu}P]V^{\mu}\right),
\label{lagrVpp}
\ee
where $P$ is the SU(3) matrix of the pseudoscalar fields. 

In a similar way, we have the Lagrangian for the coupling of vector mesons to
the baryon octet given by
\cite{Klingl:1997kf,Palomar:2002hk} \footnote{Correcting a misprint in
\cite{Klingl:1997kf}}

\be
{\cal L}_{BBV} =
\frac{g}{2}\left(tr(\bar{B}\gamma_{\mu}[V^{\mu},B]+tr(\bar{B}\gamma_{\mu}B)tr(V^{\mu})
\right),
\label{lagr82}
\ee
where $B$ is now the SU(3) matrix of the baryon octet \cite{Eck95}. Similarly,
one has also a lagrangian for the coupling of the vector mesons to the baryons
of the decuplet, which can be found in \cite{manohar}.

\begin{figure}[tb]
\begin{center}
\epsfig{file=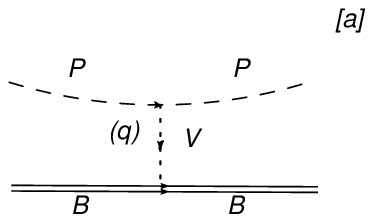, width=7cm} \epsfig{file=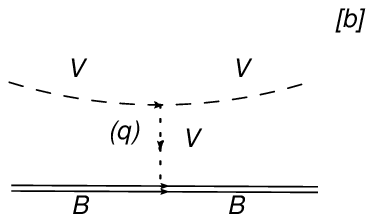, width=7cm}
\end{center}
\caption{Diagrams obtained in the effective chiral Lagrangians for interaction
of pseudoscalar [a] or vector [b] mesons with the octet or decuplet of baryons.}%
\label{f1}%
\end{figure}

With these ingredients we can construct the Feynman diagrams that lead to the $PB
\to PB$ and $VB \to VB$ interaction, by exchanging a vector meson between the
pseudoscalar or the vector meson and the baryon, as depicted in Fig.\ref{f1} .

 By looking at the 
Lagrangian of eq. (\ref{l3Vsimp})
we see that the field $V^\nu$   cannot
correspond to an external vector meson. Indeed, if this were the case, the $\nu$
index must be spatial, because $\epsilon ^0$ is zero in the limit of zero three
momentum, and then the partial derivative $\partial_\nu$  leads to
a three momentum of the vector mesons which are neglected in the approach. Then 
 $V^\nu$ corresponds to the exchanged vector
 and the analogy with the pseudoscalar and vector
interaction is much closer (see Eq. (\ref{lagrVpp}). Actually, they are formally identical 
substituting the octet of pseudoscalar fields by the octet of the vector fields,
with the additional factor $\vec{\epsilon}\vec{\epsilon }'$ in the case of the
interaction of the vector mesons. Note that $\epsilon_\mu \epsilon^\mu$ 
in Eq. (\ref{l3Vsimp}) becomes 
$-\vec{\epsilon}\vec{\epsilon }'$ and the signs of the Lagrangians also agree.

   A small amendment is in order in the case of vector mesons, which
   is due to the mixing of $\omega_8$ and the singlet of SU(3), $\omega_1$, to give the
   physical states of the $\omega$ and the $\phi$.
    In this case, all one must do is to take the
   matrix elements known for the $PB$ interaction and wherever $P$ is the
   $\eta_8$ multiply the amplitude by the factor $1/\sqrt 3$ to get the
   corresponding $\omega $ contribution and by $-\sqrt {2/3}$ to get the
   corresponding $\phi$ contribution.  Upon the approximation consistent with
   the neglect of the three momentum versus the mass of the particles (in this
   case the baryon), we can just take the $\gamma^0$ component of 
   eq. (\ref{lagr82})  and
   then the transition potential corresponding to the diagram of fig 1(b ) is
   given by
   
   \begin{equation}
V_{i j}= - C_{i j} \, \frac{1}{4 f^2} \, (k^0 + k'^0)~ \vec{\epsilon}\vec{\epsilon
} ',
\label{kernel}
\end{equation}
 where $k^0, k'^0$ are the energies of the incoming and outgoing vector mesons. 
   The same occurs in the case of the decuplet.  
    
    The $C_{i,j}$ coefficients of eq. (\ref{kernel}) can be obtained directly from 
    \cite{angels,bennhold,inoue}
    with the simple rules given above for the $\omega$ and the $\phi$, and
    substituting $\pi$ by $\rho$ and $K$ by $K^*$ in the matrix elements. The
    coefficients are obtained both in the physical basis of states or in the
    isospin basis. Here we will show results in isospin
    basis. The coefficients for the case of the decuplet can be found in 
    \cite{sarkar}.

    The next step to construct the scattering matrix is done by solving the
    coupled channels Bethe Salpeter equation in the on shell factorization approach of 
    \cite{angels,ollerulf}
   \begin{equation}
T = [1 - V \, G]^{-1}\, V,
\label{eq:Bethe}
\end{equation} 
with G the loop function of a vector meson and a baryon which we calculate in
dimensional regularization using the formula of \cite{ollerulf} and similar
values for the subtraction constants. The G function is convoluted with the 
spectral function for the vector mesons to take into account their width.

 The iteration of diagrams implicit in the Bethe Salpeter equation in the case
 of the vector mesons propagates the $\vec{\epsilon}\vec{\epsilon }'$ term 
 of the interaction, thus,
the factor $\vec{\epsilon}\vec{\epsilon }'$ appearing in the potential V,
factorizes also in the T matrix for the external vector mesons.

\section{Results} 

In this section we show results for the amplitudes obtained in the attractive
channels mentioned above.  
 Since the spin dependence only comes from the 
  $\vec{\epsilon}\vec{\epsilon }'$ factor and there is no dependence on the spin
  of the baryons, the interaction for vector-baryon states with $1/2^-$ and 
  $3/2^-$ is the same and then  we get two degenerate states each time with the
  two spins. In the case of the decuplet we get degeneracy for 
$1/2^-,3/2^-,5/2^-$. 

We summarize the results in the two Tables below

\begin{table}[ht]
      \renewcommand{\arraystretch}{1.5}
     \setlength{\tabcolsep}{0.2cm}
\begin{center}
\begin{tabular}{c|c|cc|ccccc}\hline\hline
$S,\,I$&\multicolumn{3}{c|}{Theory} & \multicolumn{5}{c}{PDG data}\\
\hline
    \vspace*{-0.3cm}
    & pole position    & \multicolumn{2}{c|}{real axis} &  &  & &  &  \\
    & {\small (convolution)}    &\multicolumn{2}{c|}{{\small
    (convolution)}} & \\
    &   & mass & width &name & $J^P$ & status & mass & width \\
    \hline
$0,1/2$ & --- & 1696  & 92  & $N(1650)$ & $1/2^-$ & $\star\star\star\star$ & 1645-1670
& 145-185\\
  &      &       &     & $N(1700)$ & $3/2^-$ & $\star\star\star$ &
	1650-1750 & 50-150\\
       & $1977 + {\rm i} 53$  & 1972  & 64  & $N(2080)$ & $3/2^-$ & $\star\star$ & $\approx 2080$
& 180-450 \\	
   &     &       &     & $N(2090)$ & $1/2^-$ & $\star$ &
 $\approx 2090$ & 100-400 \\
 \hline
$-1,0$ & $1784 + {\rm i} 4$ & 1783  & 9  & $\Lambda(1690)$ & $3/2^-$ & $\star\star\star\star$ &
1685-1695 & 50-70 \\
  &       &       &    & $\Lambda(1800)$ & $1/2^-$ & $\star\star\star$ &
1720-1850 & 200-400 \\
       & $1907 + {\rm i} 70$ & 1900  & 54  & $\Lambda(2000)$ & $?^?$ & $\star$ & $\approx 2000$
& 73-240\\
       & $2158 + {\rm i} 13$ & 2158  & 23  &  &  &  & & \\
       \hline
$-1,1$ & $ --- $ & 1830  & 42  & $\Sigma(1750)$ & $1/2^-$ & $\star\star\star$ &
1730-1800 & 60-160 \\
  & $ --- $   & 1987  & 240  & $\Sigma(1940)$ & $3/2^-$ & $\star\star\star$ & 1900-1950
& 150-300\\
   &     &       &   & $\Sigma(2000)$ & $1/2^-$ & $\star$ &
$\approx 2000$ & 100-450 \\\hline
$-2,1/2$ & $2039 + {\rm i} 67$ & 2039  & 64  & $\Xi(1950)$ & $?^?$ & $\star\star\star$ &
$1950\pm15$ & $60\pm 20$ \\
         & $2083 + {\rm i} 31 $ &  2077     & 29  &  $\Xi(2120)$ & $?^?$ & $\star$ &
$\approx 2120$ & 25  \\
 \hline\hline
    \end{tabular}
\caption{The properties of the 9 dynamically generated resonances and their possible PDG
counterparts.}
\label{tab:octet}
\end{center}
\end{table}

\begin{table*}[ht]
      \renewcommand{\arraystretch}{1.5}
     \setlength{\tabcolsep}{0.2cm}
\begin{center}
\begin{tabular}{c|l|cc|lcclc}\hline\hline
$S,\,I$&\multicolumn{3}{c|}{Theory} & \multicolumn{5}{c}{PDG data}\\\hline
        & pole position &\multicolumn{2}{c|}{real axis} & name & $J^P$ & status & mass & width \\
        &               & mass & width & \\\hline
$0,1/2$ & $1850+i5$   & 1850  & 11  & $N(2090)$ & $1/2^-$ & $\star$ & 1880-2180 & 95-414\\
        &             &       &     & $N(2080)$ & $3/2^-$ & $\star\star$ & 1804-2081 & 180-450\\ 
        &       &  $2270(bump)$ &  & $N(2200)$ & $5/2^-$ & $\star\star$ & 1900-2228 & 130-400\\	         
\hline
$0,3/2$ & $1972+i49$  & 1971  & 52  & $\De(1900)$ & $1/2^-$ & $\star\star$ & 1850-1950 & 140-240 \\	
	&             &       &     & $\De(1940)$ & $3/2^-$ & $\star$ & 1940-2057 & 198-460   \\
        &             &       &     & $\De(1930)$ & $5/2^-$ & $\star\star\star$ & 1900-2020  & 220-500   \\
	&             & $2200 (bump)$  &     & $\De(2150)$ & $1/2^-$ & $\star$ & 2050-2200  & 120-200  \\
\hline
$-1,0$  & $2052+i10$  & 2050  & 19  & $\Lambda(2000)$ & $?^?$ & $\star$  & 1935-2030 & 73-180\\
\hline
$-1,1$  & $1987+i1$   & 1985  & 10   & $\Sigma(1940)$ & $3/2^-$  & $\star\star\star$ &
1900-1950 & 150-300 \\
        & $2145+i58$  & 2144  & 57  & $\Sigma(2000)$ & $1/2^-$  & $\star$ & 1944-2004 &
	116-413\\
	& $2383+i73$  & 2370  & 99 & $\Sigma(2250)$ & $?^?$ & $\star\star\star$ & 2210-2280 &
	60-150\\
	&   &   &  & $\Sigma(2455)$ & $?^?$ & $\star\star$ & 2455$\pm$10 &
	100-140\\
\hline
$-2,1/2$ & $2214+i4$  & 2215  & 9  & $\Xi(2250)$ & $?^?$ & $\star\star$ & 2189-2295 & 30-130\\
	 & $2305+i66$ & 2308  & 66 & $\Xi(2370)$ & $?^?$ & $\star\star$ & 2356-2392 & 75-80 \\
         & $2522+i38$ & 2512  & 60  & $\Xi(2500)$ & $?^?$ & $\star$ & 2430-2505 & 59-150\\	
\hline
$-3,1$   & $2449+i7$   & 2445 & 13  & $\Omega(2470)$   & $?^?$	 & $\star\star$ & 2474$\pm$12 & 72$\pm$33\\ 
 \hline\hline
    \end{tabular}
\caption{The properties of the 10 dynamically generated resonances and their possible PDG
counterparts. We also include the $N^*$ bump around 2270 MeV and the $\Delta^*$ bump around 2200 MeV. }
\label{tab:decu}
\end{center}
\end{table*}  

As one can see in Table \ref{tab:octet} there are states which one can easily
associate to known resonances. There are ambiguities in other cases. One can also
see that in several cases the degeneracy in spin that the theory predicts is
clearly visible in the experimental data, meaning that there are 
several states with about 50 MeV 
or less mass difference
between them.  In some cases, the theory predicts quantum numbers for 
resonances which have no spin and parity associated. It would be interesting to
pursue the experimental research to test the theoretical predictions. 

The results for the decuplet, summarized in Table \ref{tab:decu}, are equally interesting.
We observe here that one gets $\Delta$ states as well as $N^*$, some of which
can be clearly identified with known resonances. It is also nice to see the spin
degeneracy present  in  $N^*$, $\Delta$ and $\Sigma$ experimental data.
  The predictions made here for resonances not observed should be a stimulus for
further search of such states. In this
sense it is worth noting the experimental program at Jefferson Lab 
\cite{Price:2004xm} to investigate the $\Xi$ resonances. We are
confident that the predictions  shown here stand on solid grounds and anticipate much
progress in the area of baryon spectroscopy and on the understanding of the
nature of the baryonic resonances. 

\section*{Acknowledgments}

This work is partly supported by DGICYT contract number
FIS2006-03438.
This research is  part of the EU Integrated Infrastructure Initiative Hadron Physics Project
under  contract number RII3-CT-2004-506078.

\end{document}